# Generation of entanglement between a highly wave-packet-tunable photon and a spin-wave memory in cold atoms

YA LI,[1,2] YAFEI WEN,[1,2,3] SHENGZHI WANG,[1,2] CHAO LIU,[1,2] HAILONG LIU,[1,2] MINJIE WANG,[1,2] CAN SUN,[1,2] YAN GAO,[1,2] SHUJING LI,[1,2,4] AND HAI WANG[1,2,5]

[1]*The State Key Laboratory of Quantum Optics and Quantum Optics Devices, Institute of Opto-Electronics, Shanxi University, Taiyuan 030006, China*
[2]*Collaborative Innovation Center of Extreme Optics, Shanxi University, Taiyuan 030006, China*
[3]*Department of Physics, Taiyuan Normal University, Jinzhong 030619,China*
[4]*lishujing@sxu.du.cn*
[5]*wanghai@sxu.edu.cn*

**Abstract:** Controls of waveforms (pulse durations) of single photons are important tasks for effectively interconnecting disparate atomic memories in hybrid quantum networks. So far, the waveform control of single photon that is entangled with an atomic memory remains unexplored. Here, we demonstrated control of waveform length of the photon that is entangled with an atomic spin-wave memory by varying light-atom interaction time in cold atoms. The Bell parameter $S$ as a function of the duration of photon pulse is measured, which shows that violations of Bell equality can be achieved for the photon pulse in the duration range from 40 ns to 50 μs, where, $S=2.64\pm0.02$ and $S=2.26\pm0.05$ for the 40-ns and 50-μs durations, respectively. The measured results show that S parameter decreases with the increase in the pulse duration. We confirm that the increase in photon noise probability per pulse with the pulse-duration is responsible for the $S$ decrease.

## 1. Introduction

Quantum networks (QWs) play an important role in distributed long-distance quantum communications [1-3] and quantum computations [4, 5]. The quantum interfaces that generate entanglement between an atomic memory and a photon [6-16] can be used as quantum node and then are basic building blocks for QWs. The elementary step in QWs is to establish entanglement between two remote memories (nodes) [1-2]. With the quantum interfaces, one can establish entanglement between two remote memories by performing two-photon Hong-Ou-Mandel interference [17-22]. Also, the remote-memory entanglement can be established by mapping quantum state of a photon entangled with an atomic memory into another remote memory [23]. Hybrid quantum networks use disparate quantum memory (QM) systems as nodes and then can benefit from the advantages of the disparate systems. In hybrid QWs, an open question is wave-packet matching of the interfacing photons that are generated from or stored in disparate memory matters (nodes). For example, the pulse widths of the single photons generated from disparate memory matters are different. Specifically, the typical pulse widths of single photons are ~3ns [24] for single quantum dots, ~0.4 [25, 26] μs for solid-state atomic ensemble, ~70-100 ns [8, 9, 12] for cold atoms, 1 ns[27] (40 μs [28]) for room-temperature atomic vapors via off-resonance (motional averaging) Duan–Lukin–Cirac–Zoller protocol, and ~0.5 μs [23] or 12 μs [29] for single atoms or ions in QED cavities. Also, for achieving high-efficiency optical quantum storages in different memory matters via various storage schemes, the required pulse-widths of the single photons are different. For example, the required pulse-width is 200ns-3μs for cold atoms [30-34] via EIT storage scheme, ~7 μs for crystals [35] via

AFC storage scheme, 0.7μs for single atom via EIT storage scheme [36], 1ns for hot atoms [37], 7ns for cold atoms [38] and ~260 fs for bulk diamond [39] via Raman storage scheme, respectively. However, HOM interference requires that the wave-packets of the two photons are perfectly overlapped, if not, visibility of HOM dip will be degraded [40]. So, the perfect wave-packet matching of the two photons is necessary for establishing hybrid QWs. Recently, two experiments have experimentally demonstrated the non-classical correlation between two different memories [41], and HOM interferometer between two photons from two different memory matters [42], where the wave-packets of the single photons are changed into appropriate values, and then the connection between the different memories are experimentally available. With various physical systems such as cavity-quantum-electrodynamics single atom [43], a single trapped ion [44], DLCZ quantum memory in cold atomic ensemble [45], cold atoms [46-49] and hot atomic vapour [50] with four-wave-mixing, the single photons with highly tunable waveform lengths have been experimentally demonstrated. However, in these experiments the single photon is only non-classically correlated with or stored in an atomic memory, it is not entangled with an atomic memory. Thus, the influences of changes of single-photon wave-packet on the atom-photon entanglement remain unexplored.

In this article, we demonstrated controls of temporal durations of waveform of the Stokes photon entangled with an atomic spin-wave memory in cold atoms via varying DLCZ-like light-atom interaction time. The retrieval efficiency $R$ and Bell parameter $S$ as a function of the width of Stokes-photon pulse are measured. The results show that in the width ranging from 40 ns to 50 μs, the retrieval efficiency is basically kept unchanged and the violation of the Bell equality can be achieved. Specifically, when the pulse width is 40 ns, we achieved $R$= 20% and $S$= 2.64±0.02, while when it is 50 μs, we achieved $R$= 15% and $S$= 2.26±0.05, respectively. To our knowledge, the 50 μs long wave packet of a single photon represents the longest wave packet of a single photon entangled with an atomic spin-wave memory. By fitting the experimental data with theoretical calculations, we reveal the reason for the decrease in $S$ parameter with the pulse duration, which results from that background noise per pulse increases with the pulse duration.

## 2. Experimental setup

The experimental setup is shown in Fig. 1(a). The atomic ensemble is a cloud of cold $^{87}$Rb atoms, whose relevant atomic levels are shown in Fig. 1(b), where, $|g\rangle = |5^2S_{1/2}, F=1\rangle$, $|s\rangle = |5^2S_{1/2}, F=2\rangle$, $|e_1\rangle = |5^2P_{1/2}, F'=1\rangle$, and $|e_2\rangle = |5^2P_{1/2}, F'=2\rangle$. Each level includes Zeeman sublevels, for example, $|g\rangle$ is written as $|g, m\rangle$, where the magnetic quantum number $m = \pm 1, 0$. After the atoms are released from the magneto-optical trap, we prepare the atoms into Zeeman levels $|g, m = \pm 1, 0\rangle$ via optical pumping with a cleaning laser. In the beginning of a spin-wave-photon entanglement (SWPE) generation trial [Fig. 1(c)], a writing pulse with a tunable duration $\tau_w$ is applied on to the atoms. The value of the duration $\tau_w$ is controlled by an AOM (see Fig.1(a)). The writing pulse is $\Delta$ =20MHz blue-detuned to $|g\rangle \to |e_2\rangle$ transition, which induces spontaneous Raman scattering of $\sigma^-$-polarized ($\sigma^+$-polarized) Stokes photons and simultaneously create the spin wave $|+\rangle$ ($|-\rangle$), where the spin wave $|+\rangle$ ($|-\rangle$) is associated with the coherence $|m_a = 0\rangle \leftrightarrow |m_b = 0\rangle$ and $|m_a = -1\rangle \leftrightarrow |m_b = -1\rangle$ ($|m_a = -1\rangle \leftrightarrow |m_b = 1\rangle$ and $|m_a = 0\rangle \leftrightarrow |m_b = 2\rangle$), which is shown in Fig. 1(b). Also, the spin coherence $|m_a = 1\rangle \leftrightarrow |m_b = 1\rangle$ is created in the Raman scattering, but is neglected in the spin wave $|+\rangle$ since it is not retrieved in the following retrieval step. In the Stokes channel (blue line in Fig. 1(a)), we transform the $\sigma^+$ ($\sigma^-$) -polarized Stokes photons into the horizontally- (*H-*) or vertically- (*V-*) polarized photon by a λ/4 plate. The joint state of the atom–photon system may be written as [7, 51] $\rho_{ap} = |0\rangle\langle 0| + \sqrt{\chi}|\Phi_{a-p}\rangle\langle\Phi_{a-p}|$, where $|0\rangle$ denotes the vacuum, $\chi(\ll 1)$ represents the probability of creating the $|\Phi_{a-p}\rangle$ state per write pulse, $\Phi_{a-p} = (\cos\vartheta|+\rangle_a|H\rangle_S + \sin\vartheta|-\rangle_a|V\rangle_S)/\sqrt{2}$ denotes the spin-wave-photon entanglement state, $|H\rangle_S$ ($|V\rangle_S$) denotes a *H-* (*V-*) polarized Stokes photon, $\cos\vartheta$ is the relevant Clebsch-Gordan coefficient with the asymmetric angle of $\vartheta \approx 0.81*(\pi/4)$ [6]. The Stokes photons are collected by a single-mode fiber labeled as $SMF_S$ in Fig. 1(b) and then have the same wave vector of $k_S$. The wave-vector of the spin-wave excitation is given by $k_{SWE} = k_W - k_S$, where, $k_W$ is the wave-vector of the write pulse. After the $SMF_S$, the Stokes photons go through a phase compensator (PC) [52], which is used to eliminate the phase shift between the *H* and *V* polarizations caused by $SMF_S$. Next, the Stokes photons pass through an optical-spectrum-filter set (OSFS) which consists of three Fabry-Perot etalons and then attenuates the write laser beam by a factor of $10^{-9}$. The OSFS transmit the Stokes field with a transmission efficiency of ~70%. Subsequently, the Stokes photon passes through a λ/2 plate. By rotating it, one can change the polarization angle $\theta_S$ of the Stokes fields. Then, the Stokes photon is guided to a polarizing beam splitter $PBS_S$, which transmits the horizontal (*H*) polarization to a detector $D_{s1}$ and reflects the vertical (*V*) polarization to a detector $D_{s2}$ for $\theta_S = 0$. The detection events at the detectors $D_{S1}$ and $D_{S2}$ are analyzed with a field programmable gate array (FPGA). As soon as a Stokes photon is detected by $D_{S1}$ or $D_{S2}$, the storage of a spin-wave in $|+\rangle$ or $|-\rangle$ mode is herald. Thus, the FPGA will send out a feed-forward signal to stop the write processes. After a storage time *t*, a $\sigma^-$-polarized read laser pulse that is resonant on the $|b\rangle \to |e_2\rangle$ transition and counter-propagates with the write beam is applied to convert the spin wave $|+\rangle$ ($|-\rangle$) into a $\sigma^-$ ($\sigma^+$)-polarized anti-Stokes photon. Constraining the phase-matching condition $k_W - k_S = k_{AS} - k_R$, where $k_R$ is the wave vector of the read laser pulse, the retrieved (anti-Stokes) photon emits into the spatial mode determined by the wave-vector $k_{AS} \approx -k_S$; i.e., it propagates in the opposite direction to the Stokes photon. The $\sigma^-$ ($\sigma^+$)-polarized anti-Stokes fields are transformed into *H* (*V*)-polarized field by a λ/4 plate labelled as $QW_{AS}$ in Fig.1(a). Thus, the atom–photon state $\Phi_{a-p}$ is transformed into the two-photon entangled state $\Phi_{pp} = \cos\vartheta|H\rangle_S|H\rangle_{AS} + \sin\vartheta|V\rangle_S|V\rangle_{AS}$, where, $|H\rangle_{AS}$ ($|V\rangle_{AS}$) denotes a *H-* (*V-*) polarized anti-Stokes (readout) photon. As shown in Fig. 1(a), the channel (red line) for collecting and detecting anti-Stokes (readout) photon is similar to that for the Stokes (write-out) photon. The two outputs of $PBS_{AS}$ are sent to single-photon detectors $D_{AS1}$ ($D_{AS2}$). The polarization angle $\theta_{AS}$ of the anti-Stokes field is changed by rotating a λ/2 plate before $PBS_{AS}$. After the retrieval, a cleaning pulse with duration of 200ns is applied to pump the atoms into the initial level $|g\rangle$. Then, the next trial starts. If no Stokes photon is detected during the write pulse, the atoms are pumped directly back into the initial level by the read and cleaning pulses. Subsequently the next trial starts. The time sequence of the above-mentioned SWPE is shown in Fig.1(c). While, in the measurement for the cross-correlation function $g^{(2)}$ defined in the following Eq. (5), we will apply a write pulse following by a cleaning laser pulse and then apply a read pulse. Such cycle is repeated by a large number in the measurement. The time sequence of the measurement for $g^{(2)}$ is shown in Fig.1(d).

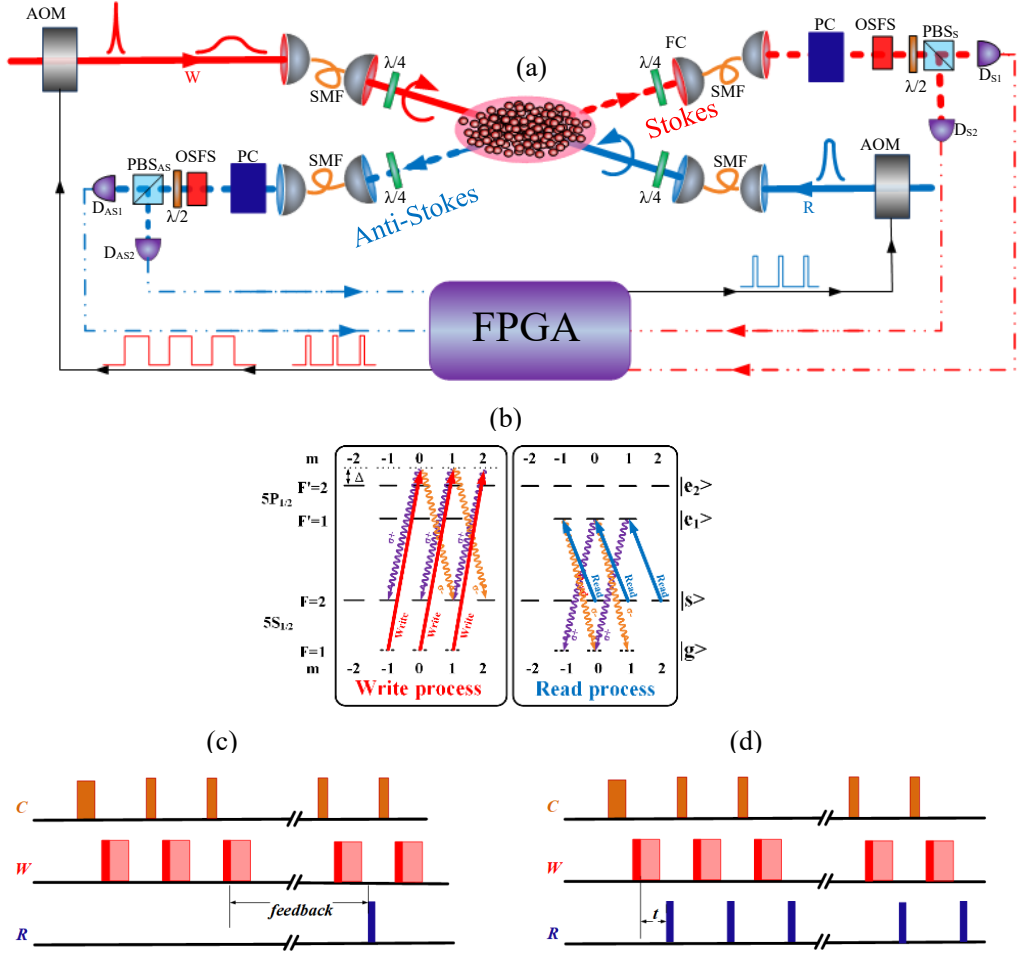

Fig. 1 Overview of the experiment. (a) Experimental setup. $\lambda/4$ ($\lambda/2$) quarter-wave (half-wave) plate; AOM: acousto-optic modulator; PC: phase compensation module; FC: fiber collimator; PBS: polarizing beam splitter; OSFS: optical-spectrum-filter set; SMF: Single-mode fiber; $D_{S1}$, $D_{S2}$, $D_{AS1}$, $D_{AS2}$: Single photon detectors (SPCM-NIR, Excelitas Technologies); FPGA: programmable gate array; (b) Relevant atomic levels. ↻(↺) represents $\sigma^-$ ($\sigma^+$) –polarized write (read) laser beam; (c) Time sequence of the measurement for measuring the Bell parameter $S$; (d) Time sequence of the experimental trials for measuring cross-correlation function $g^{(2)}$. W, C, R: write, cleaning and read laser pulses (all the lasers are DL Pro made in TOPTICA Photonics)

## 3. Experimental results

By changing the write-laser pulse duration $\tau_w$, we varied the atom-light interaction time and then demonstrated the control of wave-packet length of the Stokes photon. In the measurement for the control of wave-packet length, we set the write laser powers to be appropriate values for different wave-packet lengths and then keep the excitation probability $\chi \approx 1\%$ unchanged. The Stokes photon counts $C_S$ are measured according to $C_S = C_{D_{S1}} + C_{D_{S2}}$, where $C_{D_{S1}}$ ($C_{D_{S2}}$) is the photon counts at $D_{S1}$ ($D_{S2}$) for the polarization angle $\theta_S = 0$. Fig.2(a),(b) and (c) show the three examples of the Stokes-photon histograms when the widths of the write pulses are $\tau_w = 40\ ns$, $\tau_w = 5\ \mu s$ and $\tau_w = 50\ \mu s$. The red lines in these figures are the temporal wave shapes of the Stokes photon without any background subtraction. The blue lines represent background noise levels. The wave-shape durations of the Stokes-photon in a, b and c are ~ $40ns$, ~ $5\,\mu s$ and ~ $50\,\mu s$, respectively, which show that the durations $\tau_S$ of the Stokes-photon pulses are the same as that of the applied write-laser pulses, i.e., $\tau_S \approx \tau_w$.

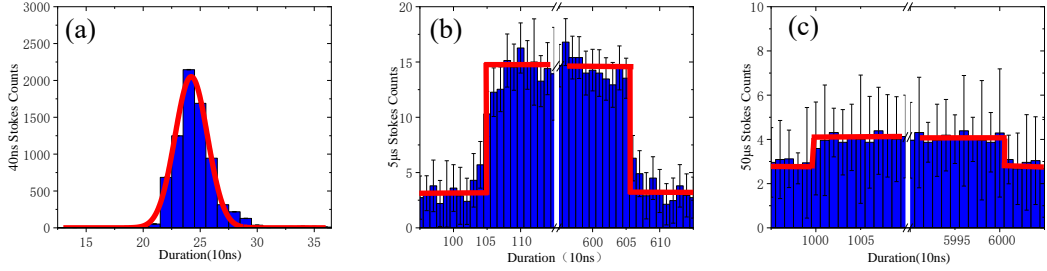

Fig. 2. Three examples of the Stokes-photon temporal wave shapes. The Stokes photons are generated by applying write pulses with (a) 40ns, (b) 5μs and (c) 50μs durations (full-width at half-maximum, FWHM). All detection events were binned into 10 ns.

The quantum correlations between the Stokes and anti-Stokes photons can be characterized by the cross-correlation function

$$g^{(2)} = P_{S,AS}/(P_S P_{AS}), \quad (1)$$

where, $P_{AS,S}$ is the coincident detection probability between the Stokes and anti-Stokes photons, $P_S$ ($P_{AS}$) denotes the probability of detecting one Stokes (anti-Stokes) photon. In our presented experiment, the coincident detection probability $P_{AS,S}$ is measured as $P_{AS,S} = P_{D_{S1},D_{AS1}} + P_{D_{S2},D_{AS2}} + P_{D_{S2},D_{AS1}} + P_{D_{S1},D_{AS2}}$ for the polarization angles $\theta_S = \theta_{AS} = 0$, where, for example, $P_{D_{S1},D_{AS1}}$ ($P_{D_{S2},D_{AS1}}$) is the probability of detecting a coincidence between the detectors $D_{S1}$ and $D_{AS1}$ ($D_{S2}$ and $D_{AS1}$). $P_S$ ($P_{AS}$) is measured as $P_S = P_{D_{S1}} + P_{D_{S2}}$ ($P_{AS} = P_{D_{AS1}} + P_{D_{AS2}}$), where, for example, $P_{D_{S1}}$ ($P_{D_{AS2}}$) is the probability of detecting a photon at $D_{S1}$ ($D_{AS2}$). In the measurements of $P_{AS,S}$, the polarization angles are set to be $\theta_S = \theta_{AS} = 0$.

The Stokes, anti-stokes and coincidence detection probabilities can be expressed as:

$$P_S = \chi \eta_S + B \eta_S \quad (2)$$
$$P_{AS} = \chi \gamma \eta_{AS} + (1-\gamma)\zeta \eta_{AS} + C \eta_{AS} \quad (3)$$
$$P_{AS,S} = \chi \gamma \eta_S \eta_{AS} + P_S P_{AS} \quad (4)$$

respectively, where, $B$ ($C$) denotes background noise probability per Stokes (anti-Stokes) pulse in the Stokes (anti-Stokes) channel, $\eta_{AS}$ ($\eta_S$) is the overall detection efficiencies in the channel, whose values are $\eta_{AS} \approx \eta_S \approx 0.3$ in the presented experiment, $\gamma$ is the retrieve efficiency, the second term in Eq. (4) represent the noise resulting from imperfect readout [53], $\xi$ is the branching ratio corresponding to the read photon transition [53].

We then measured the background noise probability $B$ per Stokes-photon (write-laser) pulse when the duration of the write pulse is changed from 40 ns to 50 μs. Since the duration of the Stokes pulse $\tau_s$ equal to that of the write-pulse duration $\tau_w$, we set the detection time interval of the Stokes photon to be equal to $\tau_w$. The noise probability $B$ includes the dark counts of the detectors and the photon counts from the write-laser leakage, but it doesn't include the emission from the atoms. Thus, we measured it without the trapped atoms. The results are shown in Fig. 3. We find that the noise probability $B$ increases linearly with the write-pulse width, i.e., it can be written as $B = k\tau_w$ with $k = 4.84 \times 10^{-3} / 100ns$.

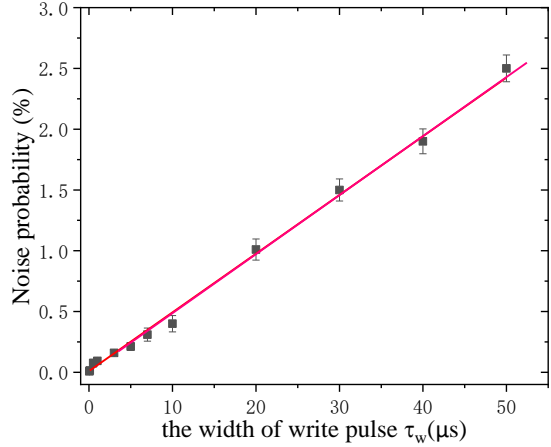

Fig. 3 The measured background noise probability B as a function of the write-pulse width, The fitting (red line) to the experimental data is based on $B = k\tau_w$ with $k = 4.84 \times 10^{-3} / 100ns$.

The retrieval efficiency is measured as $\gamma \approx (P_{AS,S} - P_S P_{AS}) / [\eta_{AS} \cdot (P_S - B\eta_S)]$. Fig. 4(a) plots the measured retrieval efficiency as a function of storage time $t$ for the case that 100-ns write pulse is applied. In this measurement, the time interval for detecting Stokes photon is selected to be 100ns. One can see that the retrieval efficiency decreases with the increase in storage time t due to spin-wave decoherence and the measured data give a 50 μs memory lifetime. Fig. 4(b) plots the measured retrieval efficiency as a function of the write-pulse width $\tau_w$ for a fixed storage time $t$ of ~1μs, where the time interval for detecting Stokes photon is set to be the width of $\tau_w$. As shown in Fig. 4(b), the measured retrieval efficiency decrease with the increase in the write-pulse width $\tau_w$. The reason for such decrease is that spin-wave decoherence rate increase with the write-pulse width $\tau_w$, where a long width $\tau_w$ leads to a long time for creating the spin wave.

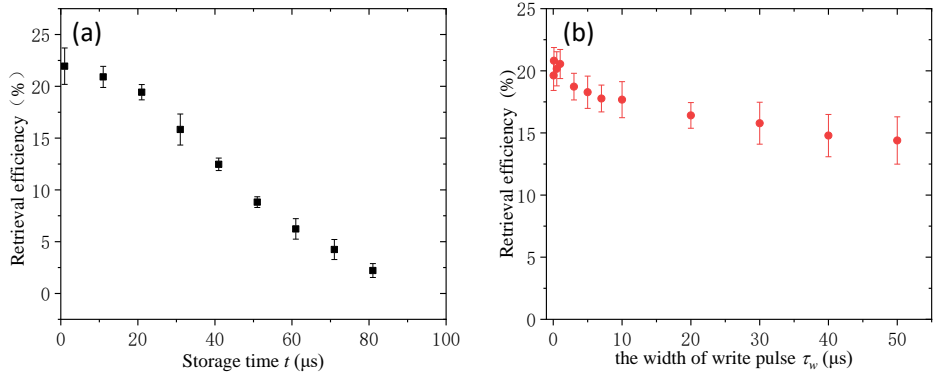

Fig. 4 Retrieval efficiency $\gamma$ as a function of storage time $t$ (a) and the write-pulse width $\tau_w$ (b) for the excitation probability $\chi=1\%$ .

According to the Eq. (1), we measured the cross-correlation function g$^{(2)}$ as a function of the write-pulse width $\tau_w$ for a fixed excitation probability $\chi \approx 1\%$ per trial (write pulse). In the measurement, the time interval for detecting Stokes photon is set to be $\tau_w$ . As shown in Fig.5, the value of g$^{(2)}$ decrease with the increase in the width $\tau_w$. We attribute such decrease to the increase in the background noise $B$ and the decrease in the retrieval efficiency $\gamma$ with the width $\tau_w$. This view point can be explained by the following theoretical calculations. According the expressions of $P_{S,AS}$, $P_S$ and $P_{AS}$, we rewrite the quantum correlation g$^{(2)}$ as:

$$g^{(2)} = P_{S,AS}/(P_S P_{AS}) = 1 + \frac{\gamma(\tau_w)}{(B+\chi)\gamma(\tau_w)+(B+\chi)(1-\gamma(\tau_w))\xi+C+BC/\chi}$$
$$= 1 + \frac{\gamma(\tau_w)}{(k\tau_w+\chi)\gamma(\tau_w)+(k\tau_w+\chi)(1-\gamma(\tau_w))\xi+C+k\tau_w C/\chi} \quad (5)$$

where, $B=k\tau_w$, which is given in Fig.3, has been put into the Eq.(5), $\gamma(\tau_w)$ is the retrieval efficiency as a function of the width $\tau_w$, whose values can be obtained from the measured data shown in Fig.4(b). The black solid curve in Fig.5 is the fitting to the experimental data based on the Eq.(5) with the adjustable parameter $\xi$. One can see that the fitting is well agreement in the experimental data g$^{(2)}$.

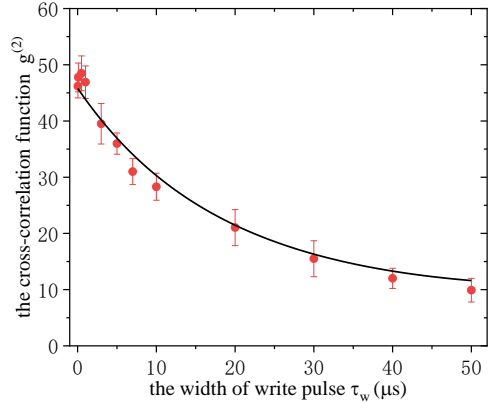

Fig.5 The measured cross-correlation function g$^{(2)}$ as a function of the write-pulse width $\tau_w$.

The black solid curve is the fitting to the measured data based on Eq. (5) for the parameters including the excitation probability $\chi=1\%$, and $\xi=0.27$. In the measurement, the time interval for detecting Stokes photon is set to be the corresponding $\tau_w$.

The quality of the spin-wave-photon entanglement can be described by the Clauser–Horne–Shimony–Holt Bell parameter $S$, which is written as

$$S = |E(\theta_S, \theta_{AS}) - E(\theta_S, \theta'_{AS}) + E(\theta'_S, \theta_{AS}) + E(\theta'_S, \theta'_{AS})| < 2$$

with the correlation function $E(\theta_S, \theta_{AS})$ defined by

$$\frac{C_{D_{S_1},D_{AS_1}}(\theta_S,\theta_{AS}) + C_{D_{S_2},D_{AS_2}}(\theta_S,\theta_{AS}) - C_{D_{S_1},D_{AS_2}}(\theta_S,\theta_{AS}) - C_{D_{S_2},D_{AS_1}}(\theta_S,\theta_{AS})}{C_{D_{S_1},D_{AS_1}}(\theta_S,\theta_{AS}) + C_{D_{S_2},D_{AS_2}}(\theta_S,\theta_{AS}) + C_{D_{S_1},D_{AS_2}}(\theta_S,\theta_{AS}) + C_{D_{S_2},D_{AS_1}}(\theta_S,\theta_{AS})}, \quad (6)$$

where, for example, $C_{D_{S_1},D_{AS_1}}(\theta_S,\theta_{AS})$ ($C_{D_{S_2},D_{AS_2}}(\theta_S,\theta_{AS})$) denotes the coincidence counts between detectors $D_{S_1}$ ($D_{S_2}$) and $D_{AS_1}$ ($D_{AS_2}$) for the polarization angles $\theta_S$ and $\theta_{AS}$. In the $S$ measurement, we used the canonical settings $\theta_S = 0°$, $\theta'_S = 45°$, $\theta_{AS} = 22.5°$, and $\theta'_{AS} = 67.5°$. To demonstrate how the changes of wave-packet length of the Stokes photon does influences the atom-photon entanglement quality, we measured Bell parameter $S$ as a function of the write-pulse width $\tau_w$ for $\chi=1\%$ and show the measured data in Fig.6. In the measurement, the time interval for detecting Stokes photon is set to be $\tau_w$. One can see that the values of $S$ decrease with the increase in the width $\tau_w$. At $\tau_W=40ns$, $S=2.64\pm0.02$, which violates the CHSH inequality by 32 standard deviations. At $\tau_w=50\mu s$, $S=2.26\pm0.05$, which violates the CHSH inequality by 5 standard deviations. The dependence of the Bell parameter on the write-pulse width $\tau_w$ can be written as

$$S(\tau_w) = 2\sqrt{2}V_0 \frac{g^{(2)}(\tau_w)-1}{g^{(2)}(\tau_w)+1} \quad (7)$$

where, $V_0$ is the initial visibility ($\tau_w=100ns$), which is $V_0 \approx 95.7\%$ in the presented experiment and is main limited to the imperfect phase compensation of the optical elements and the asymmetry angle $\vartheta$. Based on the Eq.(7), we plot the fitting (red solid curve) to the experimental data of $S$ in Fig.6, where the values of $g^{(2)}(\tau_w)$ are obtained from the data in Fig.5. One can see that the fitting is well agreement in the experimental data $S$, showing that the reduction of $S$ as the width $\tau_w$ has the same reason, i.e., the reduction of $S$ results from the increase in the background noise $B$ and the decrease in the retrieval efficiency with the width $\tau_w$.

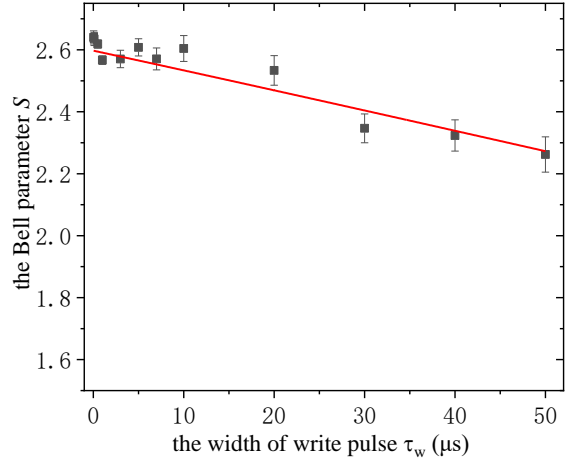

Fig.6 The measured Bell parameter *S* as a function of the write-pulse width $\tau_w$. The red solid curve is the fitting to the measured data based on Eq. (7) with the parameter $V_0 \approx 95.7\%$ and the values of $g^{(2)}(\tau_w)$ taken from the data in Fig.5.

## 4. Conclusion

Based on atom-photon entanglement generation via DLCZ scheme, we demonstrated the control of waveform length of the Stokes photon by varying light-atom interaction time in cold atoms. The Bell parameter *S* as the function of the duration of the Stokes photon pulse (write-pulse width) is measured, which shows that violations of the Bell equality can be achieved for the Stokes photon pulse in the duration ranging from 40 ns to 50 μs with *S*= 2.64$\pm$0.02 and *S*= 2.26$\pm$0.05 for 40 ns and 50 μs durations, respectively. To our knowledge, the 50 μs long wave packet of a single photon represents the longest wave packet of single photon non-classically correlated with or entangled with an atomic spin-wave memory. In the next works, the bandwidth of the Stokes-photon should be measured, which is important quantity in Hong-Ou-Mandel interference [54]. Another question in hybrid QWs is the wavelength matching of the interfacing photons from two disparate atomic memories. By using cascaded quantum frequency conversion [41], one can overcome this question. In summary, our presented work provides a road to achieve entanglement between a single photon with highly tunable wave shape and a spin-wave memory and then benefits for hybrid quantum networks.

**Funding.** Key Project of the Ministry of Science and Technology of China (Grant No. 2016YFA0301402), The National Natural Science Foundation of China (Grants: No. 12174235), Shanxi"1331 Project" Key Subjects Construction.

**Disclosures.** The authors declare no conflicts of interest.

**Data availability.** Data underlying the results presented in this paper are not publicly available at this time but may be obtained from the authors upon reasonable request.

## References

1. Sangouard N, Simon C, de Riedmatten H, and Gisin N, "Quantum repeaters based on atomic ensembles and linear optics," Reviews of Modern Physics, **83** (1), 33-80. (2011).
2. Duan L-M, Lukin M D, Cirac J I, and Zoller P, "Long-distance quantum communication with atomic ensembles and linear optics," Nature, **414** (22), 413-418. (2001).
3. Simon C, "Towards a global quantum network," Nat Photonics, **11** (11), 678-680. (2017).
4. Kimble H J, "The quantum internet," Nature, **453** (7198), 1023-1030. (2008).

5. Reiserer A, and Rempe G, "Cavity-based quantum networks with single atoms and optical photons," Reviews of Modern Physics, **87** (4), 1379-1418. (2015).
6. Matsukevich D N, Chaneliere T, Bhattacharya M, Lan S Y, Jenkins S D, Kennedy T A, and Kuzmich A, "Entanglement of a photon and a collective atomic excitation," Phys Rev Lett, **95** (4), 040405. (2005).
7. de Riedmatten H, Laurat J, Chou C W, Schomburg E W, Felinto D, and Kimble H J, "Direct measurement of decoherence for entanglement between a photon and stored atomic excitation," Phys Rev Lett, **97** (11), 113603. (2006).
8. Chen S, Chen Y A, Zhao B, Yuan Z S, Schmiedmayer J, and Pan J W, "Demonstration of a stable atom-photon entanglement source for quantum repeaters," Phys Rev Lett, **99** (18), 180505. (2007).
9. Yang S J, Wang X J, J. Li J R, Bao X H, and J. W. Pan, "Highly Retrievable Spin-Wave-Photon Entanglement Source," Phys. Rev. Lett, **114** ( 210501 ), (2015).
10. Dudin Y O, Radnaev A G, Zhao R, Blumoff J Z, Kennedy T A, and Kuzmich A, "Entanglement of light-shift compensated atomic spin waves with telecom light," Phys Rev Lett, **105** (26), 260502. (2010).
11. Wang X J, Yang S J, Sun P F, Jing B, Li J, Zhou M T, Bao X H, and Pan J W, "Cavity-Enhanced Atom-Photon Entanglement with Subsecond Lifetime," Phys Rev Lett, **126** (9), 090501. (2021).
12. Wang S-z, Wang M-j, Wen Y-f, Xu Z-x, Ma T-f, Li S-j, and Wang H, "Long-lived and multiplexed atom-photon entanglement interface with feed-forwardcontrolled readouts," Communications Physics, **4** (168), (2021).
13. Ikuta R, Kobayashi T, Kawakami T, Miki S, Yabuno M, Yamashita T, Terai H, Koashi M, Mukai T, Yamamoto T, and Imoto N, "Polarization insensitive frequency conversion for an atom-photon entanglement distribution via a telecom network," Nat Commun, **9** (1), 1997. (2018).
14. Chang W, Li C, Wu Y K, Jiang N, Zhang S, Pu Y F, Chang X Y, and Duan L M, "Long-Distance Entanglement between a Multiplexed Quantum Memory and a Telecom Photon," Physical Review X, **9** (4), (2019).
15. Saglamyurek E, Sinclair N, Jin J, Slater J A, Oblak D, Bussieres F, George M, Ricken R, Sohler W, and Tittel W, "Broadband waveguide quantum memory for entangled photons," Nature, **469** (7331), 512-515. (2011).
16. Clausen C, Usmani I, Bussieres F, Sangouard N, Afzelius M, de Riedmatten H, and Gisin N, "Quantum storage of photonic entanglement in a crystal," Nature, **469** (7331), 508-511. (2011).
17. Zhao B, Chen Z B, Chen Y A, Schmiedmayer J, and Pan J W, "Robust creation of entanglement between remote memory qubits," Phys Rev Lett, **98** (24), 240502. (2007).
18. Yuan Z S, Chen Y A, Zhao B, Chen S, Schmiedmayer J, and Pan J W, "Experimental demonstration of a BDCZ quantum repeater node," Nature, **454** (7208), 1098-1101. (2008).
19. Yu Y, Ma F, Luo X Y, Jing B, Sun P F, Fang R Z, Yang C W, Liu H, Zheng M Y, Xie X P, Zhang W J, You L X, Wang Z, Chen T Y, Zhang Q, Bao X H, and Pan J W, "Entanglement of two quantum memories via fibres over dozens of kilometres," Nature, **578** (7794), 240-245. (2020).
20. Liu X, Hu J, Li Z F, Li X, Li P Y, Liang P J, Zhou Z Q, Li C F, and Guo G C, "Heralded entanglement distribution between two absorptive quantum memories," Nature, **594** (7861), 41-45. (2021).
21. Lago-Rivera D, Grandi S, Rakonjac J V, Seri A, and de Riedmatten H, "Telecom-heralded entanglement between multimode solid-state quantum memories," Nature, **594** (7861), 37-40. (2021).
22. Hofmann J, Krug M, Ortegel N, Gerard L, Weber M, Rosenfeld W, and Weinfurter H, "Heralded entanglement between widely separated atoms," Science, **337** (6090), 72-75. (2012).
23. Ritter S, Nolleke C, Hahn C, Reiserer A, Neuzner A, Uphoff M, Mucke M, Figueroa E, Bochmann J, and Rempe G, "An elementary quantum network of single atoms in optical cavities," Nature, **484** (7393), 195-200. (2012).
24. Yu-Ming He, Yu He, Yu-Jia Wei, Dian Wu, Mete Atatu, Christian Schneider, Sven Ho M K, Chao-Yang Lu, and Pan J-W, "On-demand semiconductor single-photon source with near-unity indistinguishability," NATURE NANOTECHNOLOGY, **8** (213), (2013).
25. Laplane C, Jobez P, Etesse J, Gisin N, and Afzelius M, "Multimode and Long-Lived Quantum Correlations Between Photons and Spins in a Crystal," Phys Rev Lett, **118** (21), 210501. (2017).
26. Kutluer K, Mazzera M, and de Riedmatten H, "Solid-State Source of Nonclassical Photon Pairs with Embedded Multimode Quantum Memory," Phys Rev Lett, **118** (21), 210502. (2017).
27. Dou J-P, Yang A-L, Du M-Y, Lao D, Gao J, Qiao L-F, Li H, Pang X-L, Feng Z, Tang H, and Jin X-M, "A broadband DLCZ quantum memory in room-temperature atoms," Communications Physics, **1** (1), (2018).
28. Dideriksen K B, Schmieg R, Zugenmaier M, and Polzik E S, "Room-temperature single-photon source with near-millisecond built-in memory," Nat Commun, **12** (1), 3699. (2021).
29. Stute A, Casabone B, Brandstatter B, Friebe K, Northup T E, and Blatt R, "Quantum-state transfer from an ion to a photon," Nat Photonics, **7** (3), 219-222. (2013).
30. Shanchao Zhang, Shuyu Zhou, M. M. T. Loy, G. K. L. Wong, and Du S, "Optical storage with electromagnetically induced transparency in a dense cold atomic ensemble," OPTICS LETTERS, **36** (23), (2011).
31. Yi-Hsin Chen, Meng-Jung Lee, I-Chung Wang, Shengwang Du, Yong-Fan Chen, Ying-Cheng Chen, and Yu I A, "Coherent Optical Memory with High Storage Efficiency and Large Fractional Delay," PHYSICAL REVIEW LETTERS, **110** (083601), (2013).
32. Ya-Fen Hsiao, Pin-Ju Tsai, Hung-Shiue Chen, Sheng-Xiang Lin, Chih-Chiao Hung, Chih-Hsi Lee, Yi-Hsin Chen, Yong-Fan Chen, Ite A. Yu, and Chen Y-C, "Highly Efficient Coherent Optical Memory Based on Electromagnetically Induced Transparency," PHYSICAL REVIEW LETTERS, **120** (183602), (2018).
33. Vernaz-Gris P, Huang K, Cao M, Sheremet A S, and Laurat J, "Highly-efficient quantum memory for polarization qubits in a spatially-multiplexed cold atomic ensemble," Nat Commun, **9** (1), 363. (2018).
34. Wang Y, Li J, Zhang S, Su K, Zhou Y, Liao K, Du S, Yan H, and Zhu S-L, "Efficient quantum memory for single-

photon polarization qubits," Nat Photonics, **13** (5), 346-351. (2019).
35. Adrian Holzäpfel, Jean Etesse , KrzysztofT Kaczmarek, Alexey Tiranov, ,   N G, and Afzelius M, "Optical storage for 0.53 s in a solid-state atomic frequency comb memory using dynamical decoupling," New J. Phys. , **22** (063009), (2020).
36. Reim K F, Nunn J, Lorenz V O, Sussman B J, Lee K C, Langford N K, Jaksch D, and Walmsley I A, "Towards high-speed optical quantum memories," Nat Photonics, **4** (4), 218-221. (2010).
37. Specht H P, Nolleke C, Reiserer A, Uphoff M, Figueroa E, Ritter S, and Rempe G, "A single-atom quantum memory," Nature, **473** (7346), 190-193. (2011).
38. Ding D-S, Zhang W, Zhou Z-Y, Shi S, Shi B-S, and Guo G-C, "Raman quantum memory of photonic polarized entanglement," Nat Photonics, **9** (5), 332-338. (2015).
39. England D G, Fisher K A, MacLean J P, Bustard P J, Lausten R, Resch K J, and Sussman B J, "Storage and retrieval of THz-bandwidth single photons using a room-temperature diamond quantum memory," Phys Rev Lett, **114** (5), 053602. (2015).
40. Yuan Z S, Chen Y A, Chen S, Zhao B, Koch M, Strassel T, Zhao Y, Zhu G J, Schmiedmayer J, and Pan J W, "Synchronized independent narrow-band single photons and efficient generation of photonic entanglement," Phys Rev Lett, **98** (18), 180503. (2007).
41. Maring N, Farrera P, Kutluer K, Mazzera M, Heinze G, and de Riedmatten H, "Photonic quantum state transfer between a cold atomic gas and a crystal," Nature, **551** (7681), 485-488. (2017).
42. Craddock A N, Hannegan J, Ornelas-Huerta D P, Siverns J D, Hachtel A J, Goldschmidt E A, Porto J V, Quraishi Q, and Rolston S L, "Quantum Interference between Photons from an Atomic Ensemble and a Remote Atomic Ion," Phys Rev Lett, **123** (21), 213601. (2019).
43. Morin O, Korber M, Langenfeld S, and Rempe G, "Deterministic Shaping and Reshaping of Single-Photon Temporal Wave Functions," Phys Rev Lett, **123** (13), 133602. (2019).
44. Almendros M, Huwer J, Piro N, Rohde F, Schuck C, Hennrich M, Dubin F, and Eschner J, "Bandwidth-tunable single-photon source in an ion-trap quantum network," Phys Rev Lett, **103** (21), 213601. (2009).
45. Farrera P, Heinze G, Albrecht B, Ho M, Chavez M, Teo C, Sangouard N, and de Riedmatten H, "Generation of single photons with highly tunable wave shape from a cold atomic ensemble," Nat Commun, **7** 13556. (2016).
46. Du S, Kolchin P, Belthangady C, Yin G Y, and Harris S E, "Subnatural linewidth biphotons with controllable temporal length," Phys Rev Lett, **100** (18), 183603. (2008).
47. Liao K, Yan H, He J, Du S, Zhang Z M, and Zhu S L, "Subnatural-linewidth polarization-entangled photon pairs with controllable temporal length," Phys Rev Lett, **112** (24), 243602. (2014).
48. Zhao L, Guo X, Liu C, Sun Y, Loy M M T, and Du S, "Photon pairs with coherence time exceeding 1 µs," Optica, **1** (2), (2014).
49. Han Z, Qian P, Zhou L, Chen J F, and Zhang W, "Coherence time limit of the biphotons generated in a dense cold atom cloud," Sci Rep, **5** 9126. (2015).
50. Shu C, Chen P, Chow T K, Zhu L, Xiao Y, Loy M M, and Du S, "Subnatural-linewidth biphotons from a Doppler-broadened hot atomic vapour cell," Nat Commun, **7** 12783. (2016).
51. Wen Y, Zhou P, Xu Z, Yuan L, Zhang H, Wang S, Tian L, Li S, and Wang H, "Multiplexed spin-wave–photon entanglement source using temporal multimode memories and feedforward-controlled readout," Physical Review A, **100** (1), (2019).
52. Tian L, Xu Z, Chen L, Ge W, Yuan H, Wen Y, Wang S, Li S, and Wang H, "Spatial Multiplexing of Atom-Photon Entanglement Sources using Feedforward Control and Switching Networks," Phys Rev Lett, **119** (13), 130505. (2017).
53. Heller L, Farrera P, Heinze G, and de Riedmatten H, "Cold-Atom Temporally Multiplexed Quantum Memory with Cavity-Enhanced Noise Suppression," Physical Review Letters, **124** (21), (2020).
54. Hong C K, Ou Z Y, and Mandel L, "Measurement of subpicosecond time intervals between two photons by interference," Phys Rev Lett, **59** (18), 2044-2046. (1987).